\documentclass[iop]{emulateapj}

\usepackage{amsmath}
\usepackage{natbib}
\usepackage{graphicx}

\def\gtwid{\mathrel{\raise.3ex\hbox{$>$\kern-.75em\lower1ex\hbox{$\sim$}}}}
\def\ltwid{\mathrel{\raise.3ex\hbox{$<$\kern-.75em\lower1ex\hbox{$\sim$}}}}
\def\\{\hfil\break}

                       \def\sun{\hbox{$\odot$}}

\def\lesssim{\mathrel{\hbox{\rlap{\hbox{\lower4pt\hbox{$\sim$}}}\hbox{$<$}}}}
\def\gtrsim{\mathrel{\hbox{\rlap{\hbox{\lower4pt\hbox{$\sim$}}}\hbox{$>$}}}}

\newcommand{\mamo}[1]{\mbox{$#1$}}
\newcommand{\unit}[1]{\ifmmode \:\mbox{\rm #1}\else \mbox{#1}\fi}

\newcommand{\sbr}[1]{_{\rm #1}}

\newcommand{\ten}[1]{\mamo{\times 10^{#1}}}

\newcommand{\figref}[1]{Fig.~\ref{fig:#1}}

\newcommand{\msun}{\mamo{M_{\sun}}}

\begin{document}

\title{Dark Matter Halos in Galaxies and Globular Cluster Populations}
\author{Michael J. Hudson\altaffilmark{1}}
\affil{Dept.\ of Physics and Astronomy, University of Waterloo, Waterloo, ON N2L 3G1, Canada.}
\author{Gretchen L. Harris}
\affil{Dept.\ of Physics and Astronomy, University of Waterloo, Waterloo, ON N2L 3G1, Canada.}
\author{William E. Harris}
\affil{Dept.\ of Physics and Astronomy, McMaster University, Hamilton, ON L8S 4M1, Canada.}

\email{mjhudson@uwaterloo.ca}
\altaffiltext{1}{Perimeter Institute for Theoretical Physics, Waterloo, Ontario, Canada.}
\shorttitle{}
\shortauthors{Hudson, Harris, and Harris}

\begin{abstract}
We combine a new, comprehensive database for globular cluster populations in all types of galaxies with a new calibration of galaxy halo masses based entirely on weak lensing.
Correlating these two sets of data, we find that the mass ratio  $\eta \equiv M\sbr{GCS}/M\sbr{h}$ (total mass in globular clusters, divided by halo mass) is essentially constant at $\langle \eta \rangle \sim 4 \times 10^{-5}$, strongly confirming earlier suggestions in the literature.  Globular clusters are the only known stellar population that formed in essentially direct proportion to host galaxy halo mass.  The intrinsic scatter in $\eta$ appears to be at most 0.2 dex; we argue that some of this scatter is due to differing degrees of tidal stripping of the globular cluster systems between central and satellite galaxies. We suggest that this correlation can be understood if most globular clusters form at very early stages in galaxy evolution, largely avoiding the feedback processes that inhibited the bulk of field-star formation in their host galaxies.  The actual mean value of $\eta$ also suggests that about $1/4$ of the \emph{initial} gas mass present in protogalaxies collected into GMCs large enough to form massive, dense star clusters.  Finally, our calibration of $\langle \eta \rangle$ indicates that the halo masses of the Milky Way and M31 are $(1.2\pm0.5)\ten{12} M_{\sun}$ and $(3.9\pm1.8)\ten{12} M_{\sun}$ respectively. 
\end{abstract}


\maketitle

\label{firstpage}

\section{Introduction}
In the standard paradigm of galaxy formation, dark matter halos form and merge through gravitational instability, and within these halos gas cools and stars form. An essential ingredient of galaxy formation models is ``feedback'' which injects energy or momentum into the gas, either heating it or driving it outward. Commonly cited feeback mechanisms include supernovae and stellar winds in small galaxies, and active galactic nuclei (AGN) and infall gas heating in large galaxies \citep[see e.g.\ the review by][]{SilMam12}.  The bulk of the star formation and AGN activity, and hence presumably the bulk of the feedback, occurs at redshifts $z \sim 1$ -- $3$, corresponding to lookback times 8 -- 11.5 Gyr  \citep{BehWecCon13}.
Empirically star formation is most efficient for galaxies with halo masses in the intermediate range $10^{12}-10^{13} \msun$, for which 20-25\% of the baryons are converted into stars \citep{MarHud02,  LeaTinBun12, BehWecCon13, VelvanHoe14, HudGilCou13}.

Globular clusters (GCs) have been found in virtually every galaxy from dwarfs to brightest cluster galaxies (BCGs), excluding only the very tiniest dwarfs.  They are commonly regarded as relics of the earliest star-forming stages in their host galaxies, a view long confirmed by direct measurements of their ages \citep[see, e.g.][]{VanBroLea13}.  While in some galaxies affected by \emph{late} gas-rich mergers,  younger GCs are present \citep[][and later papers]{AshZep92}, the formation times for \emph{most} GCs belong to the redshift range $z \sim 2 - 8$ and thus precede the bulk of star formation in the Universe. 

The first GCs likely started forming within the pregalactic gas-rich and nearly pristine dwarfs, out of dense, particularly massive GMCs (giant molecular clouds) \citep[e.g.][among others]{Har94, har10}.  Growing evidence also suggests that the GCs were likely to have formed a little before the bulk of the metal-poor field-star population \citep[e.g.][]{Bla99,har02,pen08,Spi10}.  This ``head start'' on formation, plus their intrinsically  dense and massive structure, can be expected to have made them less easily influenced  by feedback processes even as protoclusters, than were the field stars within their host galaxies.

Moreover, the individual GCs in all kinds of galaxies exhibit almost identical distributions by mass,  heavy-element abundance, and King-model-type structures \citep[see][for a review]{har10}. Globular cluster \emph{systems} (GCSs; the ensembles of all the GCs in a given galaxy) thus provide observers with a direct view into a remarkable common thread of galaxy formation history in its initial stages.

Because most GCs started forming before most of the stars in the Universe, their formation may relate in a simpler way to the dark matter distribution than does the bulk of the stellar mass in galaxies. \cite{BlaTonMet97} first suggested that the total numbers of GCs in 
BCGs were directly proportional to the total,  dark-matter dominated mass of the galaxy cluster.  This suggestion was extended to host galaxies of all types especially by \cite{spi09}, \cite{geo10}, and \cite{hha13} among others.  Those studies used halo mass estimates based primarily on models and relatively small galaxy datasets, and spliced together a variety of methods to cover the entire range of galaxy masses.  Suggestions based on galaxy formation simulations that the GC population should correlate closely with the dark matter potential have also been made by \citet{KraGne2005} and \citet{MooDieMad2006}, though for somewhat different reasons.

In this paper, we take advantage of a large new comprehensive GCS database to derive $M\sbr{GCS}$, the total mass of all globular clusters in  a given galaxy. This dataset spans five orders of magnitude in galaxy stellar mass, and includes every galaxy type from dwarfs to supergiants. We combine these with a new observationally-based prescription for calculating galaxy halo mass, $M\sbr{h}$,  based entirely on weak lensing.  In agreement with the papers cited above, we find that the simple ratio $\eta \equiv M\sbr{GCS}/M\sbr{h}$  \citep{geo10} is virtually constant over the entire range, very much unlike the strongly nonlinear behavior of total \emph{stellar} mass $M_{\star}$ with $M\sbr{h}$.

An outline of this paper is as follows. After providing the background for the data in section 2, we discuss the distribution of $\eta$ in section 3.  In section 4 we discuss possible factors contributing to the observed scatter, and the implications of the relation for galaxy and GMC formation.

\section{Data}

\subsection{Globular Cluster Systems}

Observational detection of a GCS relies on wide-field imaging; the GC population is seen as a more-or-less spherical and centrally concentrated distribution of luminous star clusters around its host galaxy.  The GC distribution does, however, extend many effective radii outward into the halo -- easily to 100 kpc or more for large galaxies.  In nearby galaxies individual GCs are spotted as nearly round, semi-resolved sources; for a much more distant galaxy, the GCS will  show up as a statistical excess of starlike or nearly-starlike objects concentrated around the galaxy center.  
Since the brightest GCs reach luminosities $M_V < -12$ (corresponding to $M \sim 10^7 M_{\odot}$), they can be found in galaxies as distant as $\sim 200$ Mpc.  

Here, we use the recently published GCS catalog of \cite{hha13}, which lists an up-to-date total of 422 galaxies with useful measurements of their GC populations.  Calculation of $M\sbr{GCS}$ \citep[see][for details]{hha13} takes into account the modest increase in mean GC mass with galaxy size.  Its quoted measurement uncertainty can differ markedly from galaxy to galaxy, since it is primarily determined by the Poisson uncertainty in the raw GC number counts multiplied upward by the correction factor needed to account for the fraction of the population fainter than the limiting magnitude of the observations.  

Among several other global parameters of the galaxies, the GCS catalog also includes distance, $V-$ and $K-$band luminosities, and integrated colors. We use the latter quantities and the prescriptions of \cite{BelMcIKat03} to determine 
stellar mass-to-light ratios
in the $K-$band from $(B-V)$ colors and SDSS $u'-r'$ colors if available, and in the $V-$band from $(B-V)$ colors.  We correct the $V-$band estimates to the $K-$band by subtracting 0.11 dex. The r.m.s.\ scatter between $V-$ and $K-$band is $\sim 0.1$ dex. We take a median of all stellar mass estimates, discarding galaxies for which one of the estimates is discrepant by more than 0.5 dex.
Our final sample has 307 galaxies with stellar mass estimates and with $M\sbr{GCS}$.

\subsection{Halo Masses}

Halo masses are difficult to measure directly for individual galaxies, but weak gravitational lensing offers one way to determine mean halo mass for subsamples of the galaxy population \citep{BraBlaSma96, HudGwyDah98, ManSelKau06,  LeaTinBun12, VelvanHoe14, HudGilCou13}.  Here we use the recent results from the CFHTLenS collaboration in which \cite{HudGilCou13} binned $2\times 10^6$ foreground lens galaxies by stellar mass, color, and redshift, and measured the mean weak lensing distortion around all galaxies in a bin.  They fit a so-called ``halo'' model to the mean projected mass distribution that accounts for the stellar mass of the galaxy, its dark matter halo (assumed to follow a \cite{NavFreWhi97} density profile), and the contributions of neighboring galaxies.  Their fit also takes into account the fraction of galaxies in each bin that are not the dominant or ``central'' galaxy in the dark matter halos but are instead ``satellites''.  The quantity of interest for this paper is the halo mass (dark matter and stars), $M\sbr{h}$. Consistent with previous work, \cite{HudGilCou13} found that the stellar mass -- halo mass relation (SHMR) was non-linear. They also found that the relationship for high-mass galaxies evolves as a function of redshift from $z \sim 0.7$ to $z \sim 0.3$. In this paper, we extrapolate their SHMR to $z = 0$ using their equations 11--13 and the parameters labelled ``Default'' in their Table 5. This relation allows us to determine the mean halo mass $M\sbr{h} = M_{200} + M_{*}$ (where $M_{200}$ is defined as the dark matter halo mass within a radius in which the mean density is 200 times the critical density) for \emph{central} galaxies of a given observed stellar mass $M_{*}$.

\section{Results}

\begin{figure*}
\includegraphics[width=\textwidth]{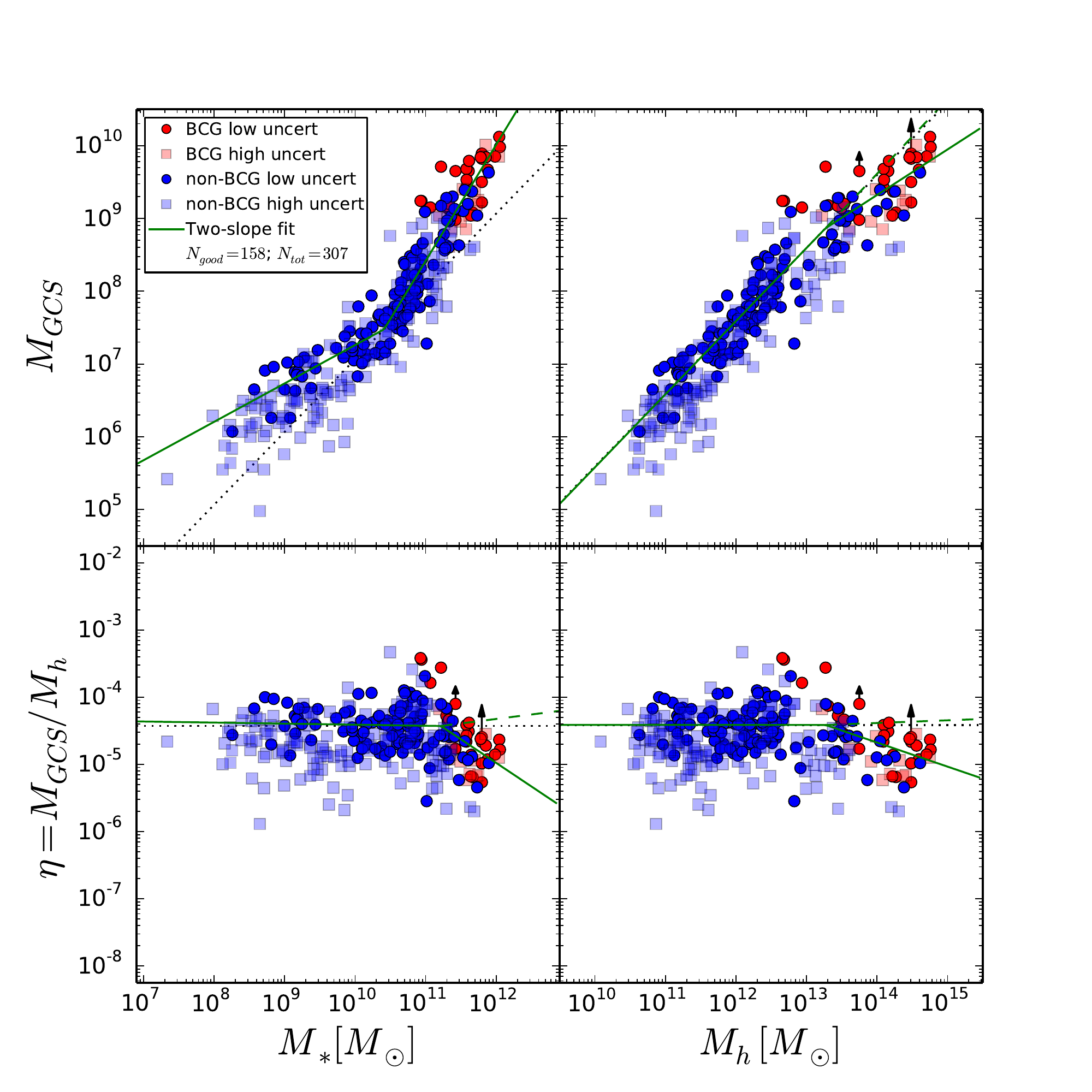}   
\caption{Total mass of all globular clusters in a galaxy ($M\sbr{GCS}$)  compared with total stellar and halo mass of the galaxy. Semi-transparent squares are for GC data with uncertainties larger than 0.1 dex, while solid circles are GC data with uncertainties $\le 0.1$ dex. Red points denote BCGs, blue points are non-BCGs (some of which may nonetheless be ``central'' galaxies in small groups). In the upper two panels the dotted lines have unit slope.  The small black arrow is attached to the point for M87 in Virgo; it shows the effect of adding all intracluster GCs in Virgo to those clustered around M87. The larger black arrow is attached to the point for NGC 4874, the central cD in the Coma cluster, and shows the same effect.   The green line is a broken power law fit to the data with small uncertainties in $M\sbr{GCS}$. The dashed green line shows the effect of correcting the high mass power law fit for IGCs, based on the observed correction in Coma.
\label{fig:gcsvshalo}}
\end{figure*}

In the upper panels of \figref{gcsvshalo}, we compare the total mass in globular clusters ($M\sbr{GCS}$) to galaxy stellar mass $M_{*}$ and halo mass $M\sbr{h}$.  While the behavior of $M_{GCS}$ versus $M_{*}$ is strongly nonlinear, it is nearly directly proportional to $M_h$.  Thus in the lower two panels, we show the mass \emph{ratio} $\eta = M\sbr{GCS}/M\sbr{h}$ versus both versions of galaxy mass.  Over most of the range of stellar (or halo) masses, the fit is consistent with a constant ratio of $\langle \eta \rangle  = (3.9\pm0.9) \times 10^{-5}$ over almost 4 orders of magnitude in stellar mass, from $\lesssim 10^8$ to $10^{12} M_{\odot}$; or almost 5 orders of magnitude in halo mass, from $10^{10}$ to $10^{15} M_{\odot}$.
For comparison, \citet{spi09}, \cite{geo10}, and \cite{hha13} obtained mean values in the range $\eta \simeq 5 - 7 \times 10^{-5}$, though with different detailed prescriptions for determining all three of $M\sbr{GCS}, M_{*}$, and $M\sbr{h}$.  Though we do not plot galaxies separately by morphological type (elliptical, spiral, S0, or irregular), notably we do not find any significant difference in $\eta$ for different galaxy types.

Our calibration of $\langle \eta \rangle$ can be used to predict the halo masses of the Milky Way and M31 from their values of $M\sbr{GCS}$.  We find $M\sbr{h, MW} = 1.2\pm0.5\ten{12} M_{\sun}$ and $M\sbr{h, M31} = 3.9\pm1.8\ten{12} M_{\sun}$, assuming 0.2 dex scatter (see below). These masses are consistent with recent estimates from abundance matching \citep{KraVikMes14} and the LG timing argument \citep{vanFarBes12, GonKraGne13}.

For $M_{*} \gtrsim 10^{11} M_{\sun}$ or $M_{h} \gtrsim 10^{13} M_{\sun}$, the ratio $\eta$ decreases slightly from a constant. Given the uncertainty in the lensing-calibrated SHMR in this range, this deviation from flatness is not statistically significant. However, a systematic trend in the data acting in the direction of the observed deviation is as follows \citep [e.g.][]{spi09}. The halo masses measured through gravitational lensing represent the entire dark matter halo around a given galaxy. Thus for a BCG-type galaxy the ``halo'' mass measured by lensing will be the mass of the entire group or cluster on scales of Mpc. In contrast, the GCS by definition is that population of GCs that are clearly clustered around the galaxy itself, regardless of whether it is a central or satellite. These $N\sbr{GC}$ totals thus usually omit any population of intracluster GCs (IGCs) which may be bound to the galaxy cluster but not any individual galaxy; for BCGs, $M\sbr{GCS}$ and thus $\eta$ will tend to be underestimated.  
We can attempt to correct for this mismatch in the accounting rules by adding the IGCs to the $N\sbr{GC}$ total for the BCG in the cluster. Counts of IGCs are generally difficult to obtain because they require especially wide-field, deep photometry, but they have been estimated usefully for two nearby clusters, Virgo and Coma. \cite{PenFerGou11} estimated that there were 47,000 IGCs within the central 520 kpc of the Coma cluster, in contrast to the 23,000 around NGC 4874 itself.  Thus the true mass of the GCS in the Coma cluster may be three times larger than the NGC 4874 GCS. 
Similarly, the IGC population found in the Virgo cluster \citep{LeeParHwa2010} may roughly double the GC total assigned to M87, the central cD.  The corrections for M87 and NGC 4874 are indicated by the arrows in Fig.~1.  In short, the evidence suggests that $\eta$ may be nearly constant over almost 5 orders of magnitude in halo mass.

\section{Discussion}

The residual scatter in $\eta$ is 0.3 dex when galaxies with the best-estimated GC counts (uncertainties $\sigma(\log M\sbr{GCS}) < 0.1$ dex) are used.  Because the halo masses $M\sbr{h}$ are not direct measurements but are inferred from the stellar mass and the mean SHMR, the scatter in the SHMR cannot be calculated directly from weak lensing, but by other methods, $\sigma(M\sbr{h})$  has been estimated at 0.15 to 0.2 dex 
\citep{BehConWec10}. 
Subtracting 0.2 dex in quadrature then suggests that the intrinsic scatter in the $M\sbr{GCS}$--$M\sbr{h}$ relation is a remarkably low $\sim 0.2$ dex.  Among several potential sources for the scatter about the mean $M\sbr{GCS}$--$M\sbr{h}$ relation, we discuss two such sources: tidal stripping of the GC system as a whole, and the red and blue GC subpopulations.

\subsection{Environmental Dependence and Tidal Stripping of the Globular Cluster System}

At large stellar or halo masses, the data are suggestive of an environmental dependence in the sense that non-BCGs have slightly lower $\eta$ than BCGs.  One obvious interpretation is that the GC system has been partially stripped in galaxies which are ``satellites'', as opposed to ``centrals'' in their host halo \citep{spi09}.  The Coma cluster may again provide a good illustration:  the two supergiants NGC 4874 and 4889 dominate the Coma center and have comparable stellar mass.  But from dynamical and X-ray studies \citep{ColDun96, ArnAghGas01}, the presence of a cD halo, and the spatial distribution of intra-cluster GCs \citep{PenFerGou11}, it is clear that NGC 4874 is currently the BCG of the Coma cluster.    At some earlier time, NGC 4889 was presumably the dominant galaxy of its own poor cluster, before it merged with a comparable-sized cluster centered on NGC 4874 to form what is now the Coma cluster.   If NGC 4889 is orbiting the potential well dominated by NGC 4874, then we expect that its own halo of dark matter and GCs has at least partially been stripped off by tidal effects, and has joined the extended Coma cluster halo. Of course the degree of stripping depends on how tightly bound a given population is. The dark matter, which is a very extended population, should be most easily stripped \citep[$\sim 65$\% retained,][]{GilHudErb13}, while the stellar light is tightly bound to the infalling galaxy and is most difficult to strip. But what is stripped would become the intracluster light. The GCS is, on average, more compact than the dark matter, but less so than the field-star light and so would be more easily stripped than the latter \cite[see][for stripping in dwarf galaxies]{SmiSanFel13}. As discussed in Section 3, the halo mass for satellites that appears in  the denominator of $\eta$ is the halo mass {\emph before} stripping, whereas the numerator is the GCS after stripping. Consequently, the $\eta-$values for the non-BCG satellites would be reduced \citep[see also][]{spi09}.

\subsection{Red vs. Blue GCs}
GCSs commonly have \emph{bimodal} distributions in GC color or metallicity \citep[e.g.][among a host of others] {lar01, pen06, har09a, mie10}; `red' GCs have mean  [Fe/H] $\simeq -0.5$, while `blue' GCs have mean [Fe/H] $\simeq -1.5$.   These two subpopulations are thought to have formed at different stages of hierarchical merging, with additional metal-poor ones accreted later from satellite dwarfs \citep[][among others]{Har94, bur01, har10, ton13}.
Direct age measurements indicate significant overlap in their age distributions, with the more metal-rich ones only $1-2$ Gyr younger than the metal-poor ones on average \citep[e.g.][] {LeaVanMen2013,han13}. 

The metal-richer clusters are almost absent in dwarfs but become progressively more prominent with galaxy luminosity, growing to roughly half the total population in giants \citep{pen06, pen08}.   However, the exact blue/red proportions differ from galaxy to galaxy even at the same luminosity, and such differences  may contribute to the scatter in Figure 1.  A direct test would be to plot $\eta$ separately for the blue and red subpopulations and see how the residual scatter changes \citep{spi09}.    Unfortunately, the numbers for $N\sbr{GCS}$ (blue, red) are not available yet for most galaxies in the catalogue of \cite{hha13}, so thorough comparisons must await future work.

\subsection{Implications for GC and galaxy formation}
The empirical fact that $M\sbr{GCS}$ is a nearly constant fraction of $M\sbr{h}$ must have its origin in early star-forming conditions.  We propose that $\eta \simeq$ constant can result if three conditions are met: 
\begin{enumerate}
\item  the \emph{initial gas mass} present in a pregalactic potential well is proportional to halo mass; 
\item  the \emph{globular cluster  formation rate} is proportional to the available gas mass; and 
\item  globular clusters form early and before  feedback effects such as stellar winds and supernovae (for dwarf galaxies), and AGN activity and infall heating (for giant galaxies) begin inhibiting star formation.  
\end{enumerate}

We suggest also that the approximate value for the ``absolute'' efficiency ratio ${\eta}$  can be understood as the product of four separate mass ratios:
\begin{align}
\eta & \sim \Bigl(\frac{M\sbr{bary}}{{M\sbr{h}}}\Bigr)\times 
\Bigl(\frac{M\sbr{GMC}}{M\sbr{bary}}\Bigr)\times 
\Bigl(\frac{M\sbr{PGC}}{M\sbr{GMC}} \Bigr)\times\Bigl(\frac{M\sbr{GC}}{M\sbr{PGC}}\Bigr) \nonumber \\
~ & \sim 0.15 \times  \Bigl(\frac{M\sbr{GMC}}{M\sbr{bary}}\Bigr) \times 0.01 \times 0.1 ~ \sim 4\times 10^{-5}
\end{align}
 Here, $M\sbr{bary}$ is the total baryonic mass in the pregalactic halo (by hypothesis, the total initial gas mass);  $M\sbr{GMC}$ is the total gas mass that collects into GMCs;  $M\sbr{PGC}$ is the gas mass within a protoglobular cluster as it begins star formation; and finally $M\sbr{GC}$ is the present-day mass of the globular cluster. The first term in equation (1) is the typical universal baryonic-to-dark mass ratio \citep{Planck13-16} and is $0.15$.  For the third term, observational evidence \citep{Har94, LadLad2003} suggests that of order 1\% of the turbulent gas in a GMC will clump into the especially dense protoclusters that will produce massive star clusters capable of surviving over the long term.  Lastly, for the fourth term, a present-day GC is typically about 10\% as massive as its initial (gaseous) protocluster because (a) the star formation efficiency within the PGC should be $\sim 0.3 - 0.5$, and (b) over 12 Gy of dynamical evolution in its host galaxy, the GC will lose 1/3 or more of its initial \emph{stellar} mass due to early rapid evolution of massive stars and later, slower tidal stripping and evaporation \citep[e.g.][]{LadLad2003,Kru2013,WebHarSil2013}.  
 
The second term $(M\sbr{GMC}/M\sbr{bary})$ is the most difficult to estimate:  it represents the fraction of all gas \emph{at the epoch of GC formation} that succeeds in cooling and collapsing into GMCs large enough to form globular clusters.
Instead we can use our direct calibration of $\langle \eta \rangle$ and invert eq.~(1) to derive
\begin{equation}
\Bigl(\frac{M\sbr{GMC}}{M\sbr{bary}}\Bigr) \sim 0.25
\end{equation}
This suggests that GMC formation in the protogalaxies was considerably more efficient than in present-day $L_{*}$ spiral galaxies for which $M\sbr{H_{2}}/M\sbr{bary}  \sim 0.01$ \citep{BosCorBoq2014}.

\subsection{Future Prospects}
Several directions for future work will lead to a better understanding of the $M\sbr{GCS}-M\sbr{h}$ correlation.  As next stages, we are searching the \cite{hha13} database for galaxies with both wide-field and multicolor data which will define $\eta$ separately for the blue and red GCs.  
Beyond that it will be necessary to make new observations, preferably for a wider range in galaxy size, type and environment. 

The correlation of $\eta$ with environment discussed in Section 4.1 suggests that tidal stripping plays an important role for satellite galaxies in the group/cluster environment. If we assume a universal $\eta$, then the observed $\eta$ could be used to measure the tidal stripping of the GCSs of individual satellite galaxies as a function of their host (group or cluster) halo mass and of their orbital properties \citep{OmaHudBeh13}.

\section*{Acknowledgements}

We acknowledge useful discussions with James Taylor.
MJH and WEH acknowledge the financial support of NSERC.

This research made use of Astropy, a community-developed core Python package for Astronomy \citep{Astropy13}.

\bibliographystyle{apj}

\label{lastpage}

\end{document}